
\input harvmac
\input amssym
\input epsf

\overfullrule=0pt
 

\def\B{{\scriptscriptstyle B}}

\def\I{{\scriptscriptstyle I}}

\def\L{{\scriptscriptstyle L}}
\def\M{{\scriptscriptstyle M}}

\def\P{{\scriptscriptstyle P}}

\def\R{{\scriptscriptstyle R}}

\def\T{{\scriptscriptstyle T}}


\def\CN{{\cal N}}

 
\def\a{\alpha}

\def\d{\delta}
\def\e{\epsilon}


\def\bdot{\dot{\beta}}


\def\half{{1/2}}
\def\quarter{{1 \over 4}}


\def\bareM{{M_0}}
\def\bareB{{B_0}}
\def\bareBbar{{\bar{B}_0}}
\def\bdot{{\dot{\beta}}}

\def\bar#1{\overline{#1}}
\def\Bbar{{\bar{B}}}
\def\BoldC{{\scriptstyle{\Bbb C}}}
\def\BoldR{{\Bbb R}}
\def\boxderiv{  {\vcenter  {\vbox  
              {\hrule height.6pt
               \hbox {\vrule width.6pt  height5pt  
                      \kern5pt 
                      \vrule width.6pt  height5pt }
               \hrule height.6pt}}} \> }
\def\BzeroNf{{{B_0}_\Nf}}
\def\BbarzeroNf{{{\bar{B_0}^\Nf}}}
\def\BbarzeroNfdagger{{{\bar{B_0}_\Nf^\dagger}}}
\def\ccdot{\hbox{\kern-.1em$\cdot$\kern-.1em}}

\def\Dbar{{\bar{D}}}

\def\fourpimu{{4\pi\mu^{\e/2}}}
\def\gammahat{\widehat{\gamma}}
\def\gtap{\raise.3ex\hbox{$>$\kern-.75em\lower1ex\hbox{$\sim$}}}

\def\L{{\scriptscriptstyle L}}
\def\Kahler{K\"ahler}
\def\KBB{{K_{\bareB\bareB^\dagger}}}
\def\KBbarBbar{{K_{\bareBbar^\dagger\bareBbar}}}
\def\KMM{{K_{\bareM^\dagger\bareM}}}

\def\KP{\kappa_\P \>}
\def\KR{\kappa_\R \>}
\def\LambdaL{{\Lambda_\L}}
\def\LLdag{{\Lambda^\dagger \Lambda}}
\def\LLLdag{{\LambdaL^\dagger \LambdaL}}

\def\ltap{\raise.3ex\hbox{$<$\kern-.75em\lower1ex\hbox{$\sim$}}}
\def\Nc{{N_c}}
\def\Nf{{N_f}}

\def\R{{\scriptscriptstyle R}}

\def\Qbar{{\bar{Q}}}

\def\sigmabar{\bar{\sigma}}

\def\therefore{{\hbox{..}\kern-.43em \raise.5ex \hbox{.}}\>\>}
\def\thetabar{{\bar{\theta}}}

\def\ZB{{Z_\B}}
\def\ZBbar{{Z_\Bbar}}
\def\ZM{{Z_\M}}

\newdimen\pmboffset
\pmboffset 0.022em
\def\oldpmb#1{\setbox0=\hbox{#1}%
 \copy0\kern-\wd0
 \kern\pmboffset\raise 1.732\pmboffset\copy0\kern-\wd0
 \kern\pmboffset\box0}


\def\appendix#1#2{\global\meqno=1\global\subsecno=0\xdef\secsym{\hbox{#1.}}
\bigbreak\bigskip\noindent{\bf Appendix. #2}\message{(#1. #2)}
\writetoca{Appendix {#1.} {#2}}\par\nobreak\medskip\nobreak}


\def\fund{  \> {\vcenter  {\vbox  
              {\hrule height.6pt
               \hbox {\vrule width.6pt  height5pt  
                      \kern5pt 
                      \vrule width.6pt  height5pt }
               \hrule height.6pt}
                         }
                   }
           \>\> }

\def\antifund{  \> \overline{ {\vcenter  {\vbox  
              {\hrule height.6pt
               \hbox {\vrule width.6pt  height5pt  
                      \kern5pt 
                      \vrule width.6pt  height5pt }
               \hrule height.6pt}
                         }
                   } }
           \>\> }

\def\sym{  \> {\vcenter  {\vbox  
              {\hrule height.6pt
               \hbox {\vrule width.6pt  height5pt  
                      \kern5pt 
                      \vrule width.6pt  height5pt 
                      \kern5pt
                      \vrule width.6pt height5pt}
               \hrule height.6pt}
                         }
              }
           \>\> }

\def\anti{ \> {\vcenter  {\vbox  
              {\hrule height.6pt
               \hbox {\vrule width.6pt  height5pt  
                      \kern5pt 
                      \vrule width.6pt  height5pt }
               \hrule height.6pt
               \hbox {\vrule width.6pt  height5pt  
                      \kern5pt 
                      \vrule width.6pt  height5pt }
               \hrule height.6pt}
                         }
              }
           \>\> }


\nref\SUSYreviews{For reviews, see K. Intriligator and N. Seiberg,
 hep-th/9509066, Nucl. Phys.  Proc. Suppl. {\bf 45BC} (1996)~1 ;
 M.E. Peskin, hep-th/9702094 ; M. Shifman, hep-th/9704114,
 Prog. Part. Nucl. Phys. {\bf 39} (1997) 1.}
\nref\SeibergI{N. Seiberg, Phys. Rev. {\bf D49} (1994) 6857.}
\nref\Pouliot{K. Intriligator and P. Pouliot, Phys. Lett. {\bf B353} (1995)
  471.}
\nref\Pesando{I. Pesando, Mod. Phys. Lett {\bf A10} 1995, 1871.}
\nref\Giddings{S.B. Giddings and J.M. Pierre, Phys. Rev. {\bf D52} (1995) 
  6065.}
\nref\Poppitz{E. Poppitz and S. Trivedi, Phys.Lett. {\bf B365} (1996) 125.}
\nref\ChoKraus{P. Cho and P. Kraus, Phys. Rev. {\bf D54} (1996) 7640.}
\nref\CSSI{C. Cs\`aki, M. Schmaltz and W. Skiba, Nucl. Phys. {\bf B487}
 (1997) 128.}
\nref\CSSII{C. Cs\`aki, M. Schmaltz and W. Skiba, Phys.Rev. {\bf D55} (1997) 
 7840.}
\nref\ILS{K. Intriligator, R.G. Leigh and N. Seiberg, Phys. Rev. {\bf D50}
 (1994) 1092.}
\nref\Intriligator{K. Intriligator, Phys. Lett. {\bf B336} (1994) 409.}
\nref\SeibergII{N. Seiberg, Nucl. Phys. {\bf B435} (1995) 129.}
\nref\GeorgiManohar{H. Georgi and A. Manohar, Nucl. Phys. {\bf B234} (1984)
  189.}
\nref\GeorgiRandall{H. Georgi and L. Randall, Nucl. Phys. {\bf B276} (1986)
  241.}
\nref\Georgi{H. Georgi, Phys. Lett. {\bf B298} (1993) 187.}
\nref\Luty{M. Luty, Phys.Rev. {\bf D57} (1998) 1531.}
\nref\Cohen{A. Cohen, D. Kaplan and A. Nelson, Phys. Lett. {\bf B412}
 (1997) 301.}
\nref\tHooft{G. t'Hooft, Nucl. Phys. {\bf B72} (1974) 461.}
\nref\Witten{E. Witten, Nucl. Phys. {\bf B160} (1979) 57.}
\nref\Grisaru{M.T. Grisaru, W. Siegel and M. Ro{\u c}ek, Nucl. Phys. {\bf B159}
 (1979) 429.}
\nref\Gates{S.J. Gates, M.T. Grisaru, M. Ro{\u c}ek and W. Siegel, {\it
 Superspace} (Benjamin/Cummings Publishing Co., Reading, MA 1983).}
\nref\tHoofttwo{G. t'Hooft, Acta Phys. Austr. Suppl. {\bf 22} (1980) 531.}


\def\LongTitle#1#2#3#4#5{\nopagenumbers\abstractfont
\hsize=\hstitle\rightline{#1}
\hsize=\hstitle\rightline{#2}
\hsize=\hstitle\rightline{#3}
\vskip 0.5in\centerline{\titlefont #4} \centerline{\titlefont #5}
\abstractfont\vskip .3in\pageno=0}
 
\LongTitle{HUTP-98/A047}{}{}
{The Confining Phase \Kahler\ Potential}
{in SUSY QCD}
{}

\centerline{Peter Cho}
\centerline{Lyman Laboratory}
\centerline{Harvard University}
\centerline{Cambridge, MA  02138}

\vskip 0.3in
\centerline{\bf Abstract}
\bigskip

	We investigate the low energy structure of the \Kahler\ potential
in SUSY QCD with $\Nf=\Nc+1$ quark flavors.  Since this theory's moduli
space is everywhere smooth, a systematic power series expansion of its
\Kahler\ potential can be developed in terms of confined meson and baryon 
fields.  Perturbation theory in the supersymmetric sigma model based upon a
momentum expansion consistent with naive dimensional analysis and $1/\Nf$
power counting exhibits some similarities with ordinary QCD chiral
perturbation theory along with several key differences.  We compute meson
and baryon wavefunction renormalization as well as \Kahler\ potential
operator mixing to leading nontrivial order.  We also deduce the asymptotic
dependence of the lowest dimension operators' coefficients upon moduli
space location along flat directions where the theory is Higgsed down to
$\Nf-1=(\Nc-1)+1$ SUSY QCD.  Although an exact form for the confining phase
\Kahler\ potential remains unknown, we find that some detailed \Kahler\
sector information can nevertheless be derived from first principles.

\Date{6/98}

\newsec{Introduction}

	During the past few years, significant progress has been made in
understanding the vacuum structure of confining $\CN=1$ supersymmetric
gauge theories \SUSYreviews.  Following Seiberg's pioneering work on SUSY
QCD \SeibergI, several examples of supersymmetric models that exist in a
confining/Higgs phase everywhere throughout moduli space have been analyzed
in detail \refs{\Pouliot{--}\CSSII}.  Massless spectra are now known in a
large class of theories, and a sizable number of dynamically generated
superpotentials which constrain gauge invariant moduli have been derived.
The often complicated superpotential expressions succinctly summarize
nonperturbative quantum deformations of classical moduli spaces.  Once one
identifies the exact superpotential in a particular theory, it is
frequently possible to deduce many others by integrating in or out
nonchiral matter fields \refs{\ILS,\Intriligator}.  An intricate web of
mutually consistent moduli space results from multiple gauge theory studies
has thus been established.

	Although the first discoveries of such nonperturbative vacua
results were impressive, they have provided only limited insight into the
sigma models which describe confining SUSY gauge theories at low energies.
Unfortunately, one can only go so far with just ground state information.
In the absence of any knowledge regarding a sigma model's kinetic sector,
it is impossible to explore a broad class of questions related to
scattering processes involving nonzero energy transfer.  To date,
relatively little has been uncovered about \Kahler\ potentials in $\CN=1$
SUSY sigma models.
 
	In this note, we initiate an investigation into the low energy
\Kahler\ sector in SUSY QCD.  We focus upon the infrared description of the
microscopic theory with $\Nf=\Nc+1$ quark flavors.  As Seiberg has
demonstrated, SUSY QCD with this particular matter content confines, and
its massless degrees of freedom consist of composite meson and baryon
superfields \SeibergI.  Unlike the superpotential whose holomorphic form is
exactly fixed by symmetry and asymptotic limit considerations, the \Kahler\
potential involves an infinite number of {\it \`a priori} undetermined
coupling constants.  However, a systematic perturbation theory can be
developed for low energy SUSY QCD in which only a finite number of \Kahler\
interaction terms contribute to scattering processes at any given order.  As
we shall see, it shares some similarities with and exhibits several key
differences from ordinary chiral perturbation theory for non-SUSY QCD.

	Our article is organized as follows.  We first construct the
leading \Kahler\ potential terms within the effective theory built upon the
vacuum at the origin of moduli space in section~2.  We next investigate
renormalization of superpotential couplings and mixing among
nonrenormalizable operators in section~3.  Deformations of $\Nf=\Nc+1$ SUSY
QCD's \Kahler\ potential along flat directions away from the origin are
explored in section~4.  Finally, we close with some thoughts on extending
our \Kahler\ sector findings in section~5, and we list our superspace
conventions in the Appendix.

\newsec{The low energy \Kahler\ potential}

	Of the many $\CN=1$ supersymmetric gauge theories which have been
studied in the past, SUSY QCD is among the simplest and best understood.  A
number of key insights into the nonperturbative dynamics of strongly
coupled supersymmetric models were first developed within this theory.  So
it represents a natural starting point for exploring confining phase
\Kahler\ potentials.

	We will focus upon SUSY QCD with $\Nf=\Nc+1 \ge 4$ quark flavors.
In the absence of any tree level superpotential, the microscopic gauge
theory has continuous symmetry group
\eqn\symgroup{G = SU(\Nc)_{\rm local} \times \bigl[ SU(\Nf)_\L \times 
SU(\Nf)_\R \times U(1)_\B \times U(1)_\R \bigr]_{\rm global},}
matter content
\eqn\matter{\eqalign{
Q^{ai} & \sim \bigl( \fund; \fund,1; 1, {1 \over \Nf} \bigr) \cr
\Qbar_{a\I} & \sim \bigl( \antifund; 1,\antifund; -1, {1 \over \Nf} \bigr)
\cr}}
and Wilsonian beta function coefficient $b_0=2\Nf-3$.  Gauge invariant meson
and baryon operators with quantum number assignments
\eqn\hadrons{\eqalign{
M^i_\I &= \Qbar Q \sim \bigl( 1; \fund, \antifund; 0, {2 \over \Nf} \bigr)
\cr
B_i &= Q^\Nc \sim \bigl(1; \antifund, 1; \Nc, {\Nc \over \Nf} \bigr) \cr
\Bbar^\I &= \Qbar^\Nc \sim \bigl(1; 1, \fund; -\Nc, {\Nc \over \Nf} \bigr)
\cr}}
act as moduli space coordinates.  Anomaly matching as well as duality
arguments compellingly demonstrate that these composite superfields
saturate the massless spectrum at the moduli space origin and that no
massive states become massless at points away from $\vev{M}=\vev{B}
=\vev{\Bbar}=0$ \refs{\SeibergI,\SeibergII}.  Since effective theory
singularities generally result from overlooked massless degrees of freedom,
$\Nf=\Nc+1$ SUSY QCD reduces at energies below its intrinsic scale
$\Lambda$ to an effective theory of mesons and baryons with a completely
smooth moduli space.

	The absence of moduli space singularities restricts the model's
dynamically generated superpotential to a simple polynomial form
\SeibergI:
\eqn\Wbare{W = \lambda_0 {\bareB \bareM \bareBbar 
- \det \bareM \over \Lambda^{2\Nf-3}}.}
We have labeled the hadrons appearing in this superpotential expression
with zero subscripts as a reminder that they represent bare fields.  The
constant $\lambda_0$ similarly represents a bare coupling.  Although it is
often absorbed into a redefinition of $\Lambda$, we shall keep track of
this dimensionless parameter for later perturbation theory purposes.

	The \Kahler\ potential in the low energy effective theory is
inherently more complicated than the superpotential, for it depends upon
both chiral and antichiral superfields as well as their derivatives.
\foot{In order for individual \Kahler\ terms to be Lorentz invariant, they
must contain an even number of both $D$ and $\Dbar$ superspace derivatives
which transform as $({1 \over 2},0)$ and $(0,{1 \over 2})$ under
$SL(2,\BoldC )$.}
Although dimensional analysis and symmetry considerations are not
sufficiently powerful to completely fix the nonholomorphic \Kahler\
potential as they essentially do for the holomorphic superpotential, they
provide useful constraints on its form:
\eqn\genKahlerformII{K = \LLdag * k\bigl({\bareM^\dagger \bareM 
\over (\LLdag)^2}, {\bareB \bareB^\dagger \over (\LLdag)^\Nc}, 
{\bareBbar^\dagger \bareBbar \over (\LLdag)^\Nc}, \cdots
\bigr).}
Since the moduli space is smooth, we can decompose the hadron fields into
quantum fluctuations about classical expectation values
\eqn\fielddecomp{
\bareM  = \vev{\bareM} + \d \bareM  \qquad
\bareB  = \vev{\bareB} + \d \bareB \qquad
\bareBbar = \vev{\bareBbar} + \d \bareBbar}
and then Taylor expand $K$ about the vacuum point
$(\vev{\bareM},\vev{\bareB}, \vev{\bareBbar})$:
\foot{We drop constant and linear terms in (2.7) which vanish when 
integrated against $\int d^4 \theta$.  We also respectively regard $\d
\bareB$ and $\d \bareBbar$ as row and column vectors in flavor space.
$\d\bareB \d\bareB^\dagger$ and $\d\bareBbar^\dagger \d\bareBbar$
consequently represent $1 \times 1$ matrices.}
\eqn\Taylorexpansion{
K=\KMM \Tr (\d\bareM^\dagger \d\bareM) + 
\KBB \d\bareB \d\bareB^\dagger + 
\KBbarBbar \d\bareBbar^\dagger \d\bareBbar + \cdots}
where $K_{H^\dagger H} \equiv \partial^2 K / \partial H^\dagger \partial H$.
This power series need not have an infinite radius of convergence.  But it
should be free of singularities for all finite, complex values of the
hadron fields.

	In order to develop a systematic perturbation theory within the low
energy sigma model, it is convenient to work with renormalized meson and
baryon fields which have unit mass dimension rather than their bare
counterparts.  We relate bare and renormalized fields as follows:
\eqn\renhadrondefns{\eqalign{
\bigl[ \LLdag \KMM \bigr]^\half \bareM &= \Lambda \ZM^\half (\mu) M(\mu) \cr
\bigl[(\LLdag)^{\Nc-1} \KBB \bigr]^\half \bareB 
  &= \Lambda^{\Nc-1} \ZB^\half (\mu) B(\mu) \cr
\bigl[(\LLdag)^{\Nc-1} \KBbarBbar \bigr]^\half \bareBbar 
  &= \Lambda^{\Nc-1} \ZBbar^\half (\mu) \Bbar(\mu). \cr}}
Since the effective theory does not contain any states with negative norm,
the Lehmann-K\"allen spectral decomposition guarantees that the
nonperturbative bounds $0 \le Z_{\M,\B,\bar{\B}}(\mu)$ $\le 1$ are
satisfied.  As we shall see, the logarithmic running to zero of the
wavefunction renormalization constants as $\mu \to 0$ forms the basis for
perturbation theory in low energy SUSY QCD.

	Working with the renormalized hadron fields in $d=4-\e$ spacetime
dimensions, we can now construct the leading independent terms in the
\Kahler\ potential's expansion about $\vev{M} = \vev{B} = \vev{\Bbar} =0$:
\foot{Following superspace conventions (A.1) and (A.2) listed in the
appendix, we include a $-1/4$ prefactor with every $D^2$ and $\Dbar^2$
which enters into the \Kahler\ potential.}
\eqn\Kfinal{\eqalign{
K &= \ZM \Tr(M^\dagger M) + \ZB (B B^\dagger + \Bbar^\dagger \Bbar) \cr &
\quad + {1 \over \LLdag} \Bigl\{ Z_1 \Tr (M^\dagger \boxderiv M) + Z_2 (
B \,\boxderiv B^\dagger + \Bbar^\dagger \boxderiv \Bbar) \cr &\quad\qquad +
(\fourpimu \lambda) \Bigl[ {Y_1 \over \sqrt{\Nf}} B ({-D^2 \over 4} M)
\Bbar + {Y_2 \over \sqrt{\Nf}} \Bigl( B M ({-D^2 \over 4} \Bbar) + ({-D^2
\over 4} B) M \Bbar \Bigr) \Bigr] \cr &\quad\qquad 
+ (\fourpimu \lambda)^*
\Bigl[ {Y_1 \over \sqrt{\Nf}} \Bbar^\dagger ({-\Dbar^2 \over 4} M^\dagger)
B^\dagger + {Y_2 \over \sqrt{\Nf}} \Bigl( \Bbar^\dagger M^\dagger
({-\Dbar^2 \over 4} B^\dagger) + ({-\Dbar^2 \over 4} \Bbar^\dagger)
M^\dagger B^\dagger \Bigr) \Bigr] \cr &\quad\qquad 
+ |\fourpimu \lambda|^2 
\Bigl[ {X_1 \over \Nf^2} \Tr (M^\dagger M) \Tr (M^\dagger M) +
{X_2 \over \Nf} \Tr (M^\dagger M M^\dagger M) \cr &
\qquad\qquad\qquad\qquad + {X_3 \over \Nf} 
(B B^\dagger) (\Bbar^\dagger \Bbar) +
{X_4 \over \Nf}\Bigl( (B B^\dagger )^2 + (\Bbar^\dagger \Bbar)^2 \Bigr) \cr 
& \qquad\qquad\qquad\qquad + {X_5 \over \Nf^2}
\Tr(M^\dagger M) (B B^\dagger + \Bbar^\dagger \Bbar) + {X_6 \over \Nf} (B M
M^\dagger B^\dagger + \Bbar^\dagger M^\dagger M \Bbar) \Bigr] \Bigr\} \cr &
\quad + O\Bigl( \Bigl({1 \over \LLdag}\Bigr)^2 \Bigr). \cr}}
We also rewrite the superpotential as
\eqn\Wfinal{W = {\fourpimu \lambda \over \sqrt{\Nf}} BM\Bbar
- {(\fourpimu \lambda)^{\Nf-2} \over \sqrt{(\Nf-1)!}} 
  {A \over \Lambda^{\Nf-3}} \det M }
where the renormalized couplings are related to the bare $\lambda_0$
parameter as
\eqn\Wcouplings{\eqalign{
{\fourpimu \lambda(\mu) \over \sqrt{\Nf}}
   &= {\ZB(\mu) \over \KBB} \Bigl[{\ZM(\mu) \over \KMM}\Bigr]^\half 
 {\lambda_0 \over (\LLdag)^{\Nf-3/2}} \cr
{\bigl(\fourpimu \lambda(\mu)\bigr)^{\Nf-2} \over \sqrt{(\Nf-1)!}} A(\mu) 
&= \Bigl[{\ZM(\mu) \over \KMM} \Bigr]^{\Nf/2} {\lambda_0 \over
   (\LLdag)^{\Nf/2}}. \cr}}
Several points about these expressions should be noted:

\item{$i)$}  SUSY QCD's discrete charge conjugation symmetry is preserved
at the origin of moduli space.  As a result, $K$ and $W$ remain invariant
under $B \leftrightarrow \Bbar^\T$ and $M \leftrightarrow M^\T$.  This
$Z_2$ reflection ensures baryon and antibaryon renormalization are
identical.  It also implies that all trilinear \Kahler\ potential operators
involving two superderivatives can be obtained via integration by parts
from the $Y_1$ and $Y_2$ terms in \Kfinal.  For example,
\eqn\trilinearKterms{(DB)(DM) \Bbar + B (DM)(D\Bbar) \longleftrightarrow  
- B (D^2M ) \Bbar.}

\item{$ii)$}  All interaction terms in $K$ and $W$ become weak
at low energies.  It is important to recall that the non-asymptotically
free sigma model flows to a free field theory in the far infrared.
Perturbative computations of quantum corrections within a systematic
momentum expansion therefore become arbitrarily accurate at lower and lower
energies.  Although nonperturbative effects are critically important in
generating the sigma model from the underlying microscopic gauge theory,
they should be small at energies below the SUSY QCD scale in the effective
theory itself.

\item{$iii)$} The dimensionless momentum expansion parameter is essentially
given by
\eqn\expansionparam{|\lambda(p^2)|^2 \quad {\buildrel p^2 \to 0 \over
\longrightarrow} \quad {{\rm constant} \over \log^2(\LLdag/p^2)}.}
We have factored out various powers of $\lambda$ from the 
coefficients of nonrenormalizable operators in $K$ and $W$ so that
$|\lambda|^2$ acts as a loop counting parameter.  All contributions to
squared scattering amplitudes from Feynman graphs with $E$ external lines
and $L$ loops are proportional to $|\lambda|^{2(E+2L-2)}$.  Unlike chiral
perturbation theory for ordinary QCD, the dominant renormalizable $BM\Bbar$
superpotential interaction does not involve derivatives.  So $|\lambda|^2$
vanishes only logarithmically as $p^2 \to 0$.  In contrast, the
$p^2/|\Lambda|^2$ expansion parameter in the QCD chiral lagrangian vanishes
linearly.

\item{$iv)$}  The sigma model becomes strongly coupled and breaks down at 
energies close to the SUSY QCD scale.  Following the rules of naive
dimensional analysis \refs{\GeorgiManohar{--}\Cohen}, we have inserted
factors of $4\pi$ into \Kfinal\ and \Wfinal\ so that all tree level and
loop contributions to individual scattering processes become comparable in
magnitude when $p^2 \simeq |\Lambda|^2$.  The dimensionless couplings in
$K$ and $W$ are then expected to be $O((4\pi)^0)$ rather than $O(4\pi)$ or
$O((4\pi)^{-1})$.

\item{$v)$}  If the number of quark flavors is large, another systematic
expansion can be formulated in the sigma model based upon $1/\Nf$.  We have
factored out various powers of $\Nf$ from the dimensionless couplings so
that Feynman graph contributions to Green's functions remain finite as $\Nf
\to \infty$.  As a result, multiple iterations of weak interactions at low
energies do not add together to mimic genuine strong effects at energies
comparable to $\Lambda$.  Since the baryon and antibaryon transform
according to fundamental and antifundamental irreps under the $SU(\Nf)_\L
\times SU(\Nf)_\R$ chiral symmetry group whereas the meson transforms
according to a 2-index bifundamental representation, large $\Nf$ power
counting within the supersymmetric sigma model is qualitatively similar to
large $\Nc$ counting in ordinary QCD \refs{\tHooft,\Witten}.

\bigskip\medskip

\topinsert
\epsfysize=12 cm\epsfbox[-50 300  150 770]{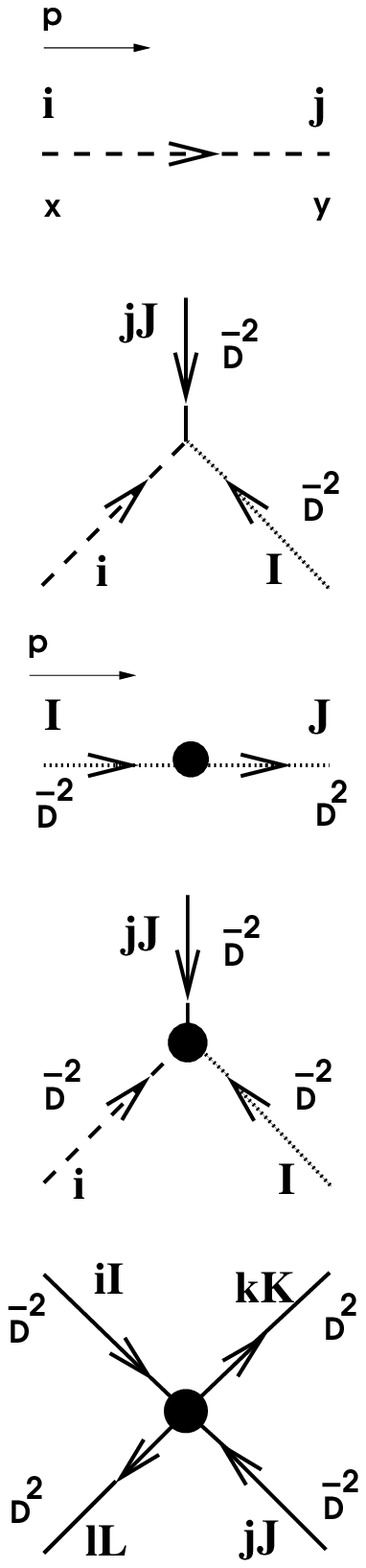}
  \hbox{\hskip 0.5truecm}
\itemitem{Fig. 1.} Supergraph Feynman rules for some representative
propagators and interaction vertices.  Solid, dashed and dotted lines
respectively denote mesons, baryons and antibaryons, while dark circles
represent individual nonrenormalizable \Kahler\ potential operators.
Arrows indicate chirality flow.
\vskip -17.3 truecm
\hskip 5.5 cm $= \quad\qquad \displaystyle{i \over p^2} 
\d^{(4)}(\theta_x -\theta_y) \d^j_i $
\vskip 1.8 truecm
\hskip 5.5 cm $ = \quad\qquad \displaystyle{i \fourpimu \lambda \over 
\sqrt{\Nf}} \d^j_i \d^I_J $
\vskip 1.4 truecm
\hskip 5.5 cm $= \quad\qquad -\displaystyle{i Z_2 p^2 \over 
\Lambda^\dagger \Lambda}\d^J_I  $
\vskip 1.5 truecm
\hskip 5.5 cm $= \quad\qquad \displaystyle{i \fourpimu \lambda \over \sqrt{\Nf}
 \Lambda^\dagger \Lambda} \cdot {Y_1 D^2_\M \over -4} \d^j_i \d^I_J $
\vskip 2.4 truecm
\hskip 5.5 cm $= \quad\qquad {2i \displaystyle{ |\fourpimu \lambda|^2 X_1 
\over \Nf^2 \Lambda^\dagger \Lambda}} 
\bigl[ \d^i_k \d^j_l \d^K_I \d^L_J + \d^i_l
\d^j_k \d^L_I \d^K_J \bigr]$
\vskip 4.5 cm
\endinsert

	With the \Kahler\ potential and superpotential expressions in hand,
we can straightforwardly work out supergraph Feynman rules for hadron
propagators and interaction vertices.  Some representative examples are
displayed in figure~1.  As can be seen in the figure, we treat the $Z_1$
and $Z_2$ terms as quadratic perturbations and do not resum these
nonrenormalizable quadratic operators into the meson and baryon
propagators.  One could choose to eliminate such noncanonical terms from
$K$ via a field redefinition.  But the hadron propagators would then
develop ghost poles and become much more complicated.  We should also note
that we have annotated the Feynman rule vertices with $D^2$ and $\Dbar^2$
symbols.  These serve as reminders that each internal chiral (antichiral)
line attached to a \Kahler\ potential vertex in a supergraph is accompanied
by a factor of $-\Dbar^2/4$ $(-D^2/4)$ coming from chiral (antichiral)
superfield functional differentiation as indicated in (A.9) [(A.10)].  A
similar rule holds for the superpotential interaction terms, except one
such squared superderivative factor is used to convert a $\int d^2 \theta$
or $\int d^2 \thetabar$ integral into an integration over all of
superspace.  After Grassmann derivatives and delta functions are
manipulated according to the rules of supergraph perturbation theory
\refs{\Grisaru,\Gates}, the integration over fermionic variables always
reduces to a single $\int d^4
\theta$ integral.  The remaining evaluation of a sigma model supergraph
then proceeds along the same lines as for any Feynman diagram in a
nonsupersymmetric bosonic field theory.

\topinsert
\epsfysize=12 cm\epsfbox[80 300 450 750]{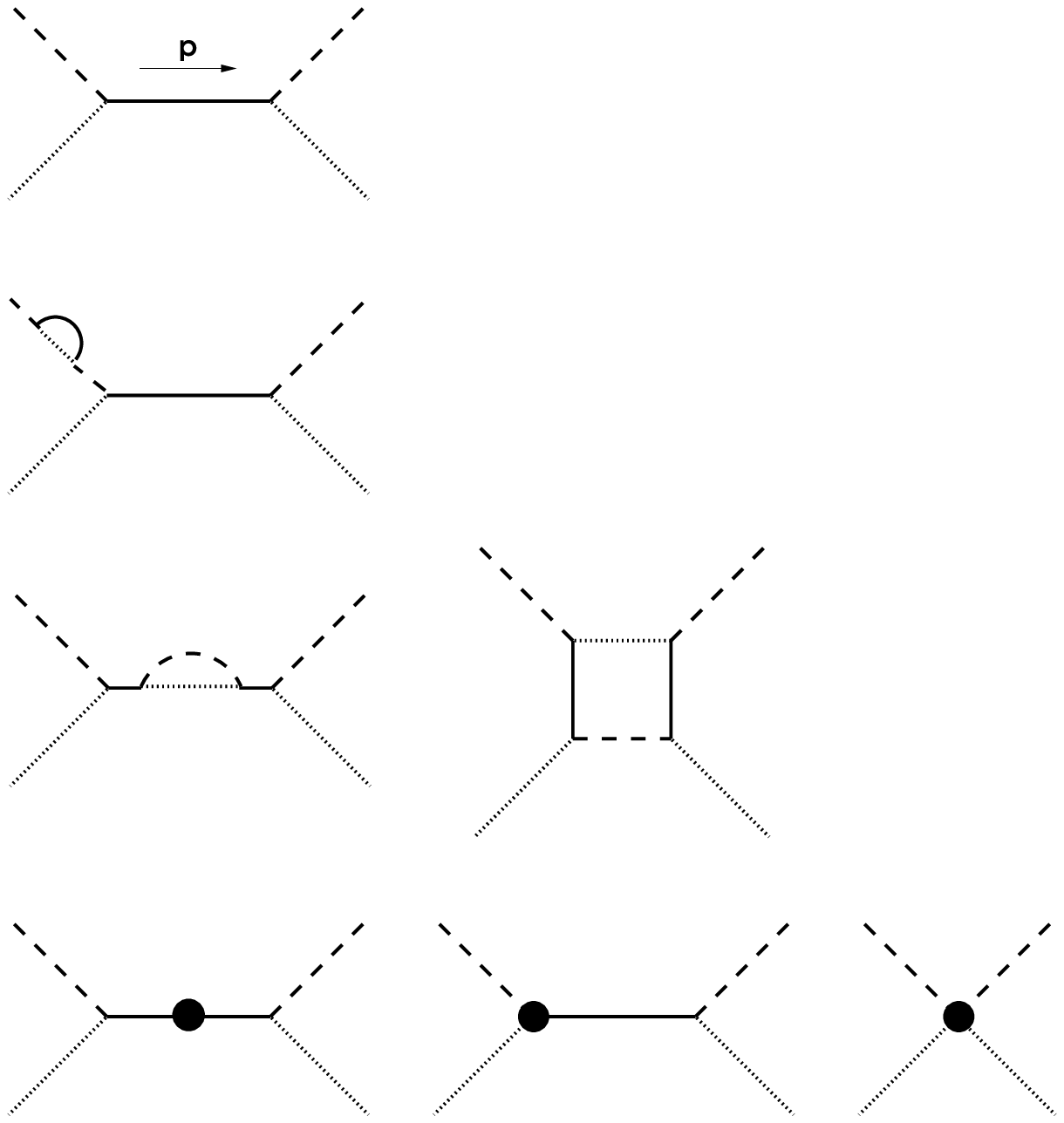}
  \hbox{\hskip 0.5truecm}
\itemitem{Fig. 2.} Order of magnitude estimates for various low order
supergraphs which mediate baryon-antibaryon scattering.  
\vskip -15.5 truecm
$$ \hskip 9 cm \sim \qquad {(4 \pi)^2 |\lambda|^2 \over \Nf p^2} $$
\vskip 1 truecm
$$ \hskip 9 cm \sim \qquad {(4 \pi)^2 |\lambda|^4 \over \Nf p^2} $$
\vskip 0.7 truecm
$$ \hskip 9 cm \sim \qquad {(4 \pi)^2 |\lambda|^4 \over \Nf^2 p^2} $$
\vskip 1 truecm
$$ \hskip 9 cm \sim \qquad {(4 \pi)^2 |\lambda|^2 \over \Nf 
 \Lambda^\dagger \Lambda } $$
\vskip 3 cm
\endinsert

	As a check on the factors of $4\pi$, $\lambda$ and $\Nf$ appearing
in \Kfinal\ and \Wfinal, we estimate the magnitudes of several supergraphs
which mediate baryon-antibaryon scattering in figure~2.  When the energy
transfer $p^2$ is much smaller than $|\Lambda|^2$, the first tree diagram
involving only the trilinear superpotential interaction dominates over the
other renormalizable loop and nonrenormalizable tree graphs.  On the other
hand, if $p^2 \simeq |\Lambda|^2$ and the number of quark flavors is small,
the coupling $\lambda$ is expected to be of order unity, and all
contributions to $B\Bbar \to B \Bbar$ scattering are comparable in size.
Finally, we observe that the tree diagrams in the figure with
nonrenormalizable operator insertions become more important than some loop
graphs involving only the renormalizable superpotential interaction as $\Nf
\to \infty$.

	It would be interesting to consider correlated combinations of the
momentum and $1/\Nf$ expansions which could be used to isolate certain
subclasses of diagrams that contribute to a particular scattering process.
However, we first need to determine how the ``fine structure constant''
$|\lambda|^2$ varies with energy.  We therefore turn to consider coupling
constant evolution in the following section.

\newsec{Renormalization}

	The low energy effective description of SUSY QCD with $\Nf=\Nc+1$
flavors is basically a massless Wess-Zumino model with an infinite number
of nonrenormalizable \Kahler\ potential operators.  Although we do not know
the precise numerical values for these operators' Wilson coefficients at
the scale around $\Lambda$ where the microscopic gauge theory matches onto
the sigma model, we can at least calculate how they evolve with energy
under the action of the renormalization group.  In the absence of mass
terms, operators of a given dimension cannot mix down into other operators
of lower dimension.  As a result, composite operator mixing involves only a
finite number of \Kahler\ potential terms at any fixed order in
$|\lambda|^2$.  This simple but important point is illustrated in figure~3
where we schematically display tree-level and one-loop contributions to
certain 1PI Green's functions.

\topinsert
\epsfysize=11 cm\epsfbox[30 180 572 630]{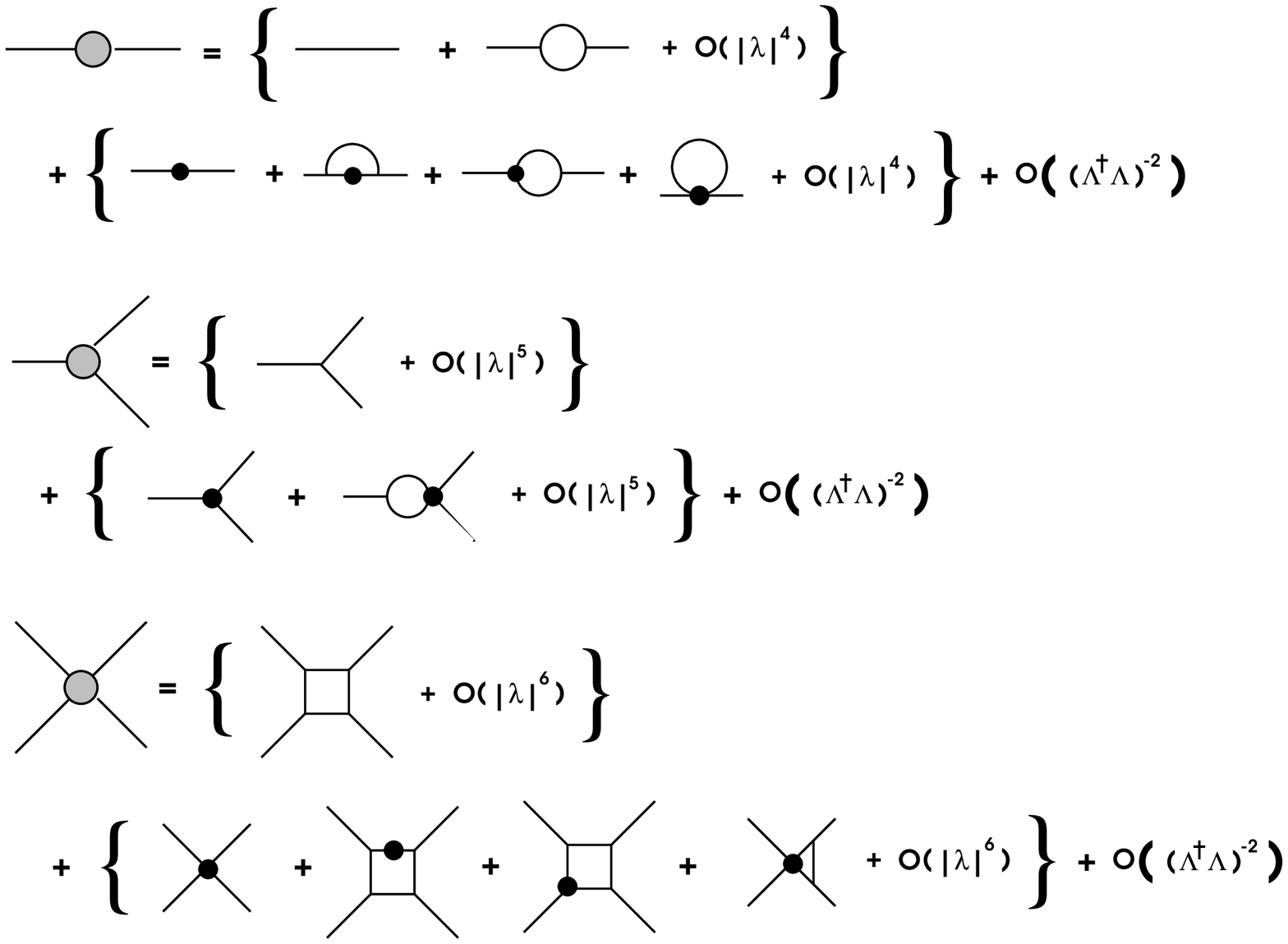}
  \hbox{\hskip 0.5truecm}
\itemitem{Fig. 3.} Some representative contributions to 1PI Green's
functions.  Mesons, baryons and antibaryons are all denoted by solid lines
in this figure.  Curly brackets enclose supergraphs which are of the same
order in $(\Lambda^\dagger \Lambda)^{-1}$.
\medskip
\endinsert

	Wavefunction renormalization in the effective theory is especially
simple.  Dimensional analysis ensures that the values of the meson and
baryon wavefunction renormalization constants are completely insensitive to
the nonrenormalizable terms in $K$.  The $\det M$ superpotential
interaction similarly has no impact.  The values for $\ZM$ and $\ZB$ can
therefore be systematically calculated order by order in $|\lambda|^2$
exactly as in a Wess-Zumino model.  After adopting the mass independent
renormalization scheme of dimensional regularization plus modified minimal
subtraction and evaluating the first one-loop supergraph displayed in
figure~3, we find the divergent meson and baryon wavefunction
renormalization constants
\eqn\wavefunctionconsts{\eqalign{
\ZM &= 1 - { |\lambda|^2 \over \Nf} \Delta + O(|\lambda|^4) \cr
\ZB &= 1 - |\lambda|^2 \Delta + O(|\lambda|^4) \cr}}
where $\Delta = 2/\e$.  We may trade $2/\e$ for its ultraviolet cutoff
regulator analog $\log(\Lambda_{\rm UV}/\mu)^2$ and insert $\ZM$ and $\ZB$
into \Wcouplings\ in order to explicitly verify that the superpotential
couplings logarithmically vanish as $\mu \to 0$:
\eqn\renWcouplings{\eqalign{
& \lambda(\mu) = {\lambda(|\Lambda|) \over 1 + \Bigl(1+\displaystyle{1
\over 2\Nf} \Bigr) |\lambda|^2 \log\Bigl(
\displaystyle{|\Lambda| \over \mu}\Bigr)^2}
\cr
& \cr
& \lambda(\mu)^{\Nf-2} A(\mu) = {\lambda
 (|\Lambda|)^{\Nf-2} A(|\Lambda|) \over 1 + {1
 \over 2} |\lambda|^2
\log\Bigl(\displaystyle{|\Lambda| \over \mu}\Bigr)^2}. \cr}}
Perturbation theory based upon $|\lambda|^2$ is consequently sensible
so long as $\mu \ll |\Lambda|$.

	Like all the other dimensionless Wilson coefficients in the
\Kahler\ potential, the wavefunction renormalization constants are real.
So it is no surprise that they develop nonholomorphic dependence upon
$|\lambda|^2$ at one-loop order. This nonanalytic behavior feeds into the
superpotential couplings.  We recall that $\lambda(\mu)$ and
$\lambda(\mu)^{\Nf-2} A(\mu)$ come from a common bare constant which
multiplies the entire superpotential for $\Nf=\Nc+1$ SUSY QCD.  But as
\renWcouplings\ demonstrates, these renormalized couplings do not run
at the same rate.  Therefore, the relative coefficient between the
$BM\Bbar$ and $\det M$ terms changes with energy scale, and the exact form
of Seiberg's superpotential is lost once one works with nonholomorphic
renormalized fields.

	We next consider mixing among the dimension-4 operators in the
\Kahler\ potential.  We assign them the names
\eqn\OPoperators{
\eqalign{O_1 &= \Tr (M^\dagger \boxderiv M) \cr
O_2 &= B \,\boxderiv B^\dagger \cr
\bar{O_2} &= \Bbar^\dagger \boxderiv \Bbar \cr}
\qquad\qquad
\eqalign{P_1 &= \KP B (D^2 M) \Bbar \cr
P_2 &= \KP  (D^2 B) M \Bbar \cr
\bar{P_2} &= \KP B M (D^2  \Bbar) \cr}}
\eqn\Roperators{\eqalign{
R_1 &= \KR [\Tr (M^\dagger M)]^2 \cr
R_4 &= \Nf \KR (B B^\dagger)^2 \cr
\bar{R_4} &= \Nf \KR (\Bbar^\dagger \Bbar)^2 \cr}
\qquad
\eqalign{
R_2 &= \Nf \KR \Tr (M^\dagger M M^\dagger M) \cr
R_5 &= \KR \Tr(M^\dagger M) B B^\dagger \cr
\bar{R_5} &= \KR \Tr(M^\dagger M) {\Bbar^\dagger \Bbar} \cr}
\qquad
\eqalign{
R_3 &= \Nf \KR (B B^\dagger)(\Bbar^\dagger \Bbar) \cr
R_6 &= \Nf \KR B M M^\dagger B^\dagger \cr
\bar{R_6} &= \Nf \KR \Bbar^\dagger M^\dagger M \Bbar \cr}}
with $\kappa_\P = -(\fourpimu \lambda)/(4 \sqrt{\Nf})$ and $\kappa_\R =
|\fourpimu \lambda |^2 / \Nf^2$.  These operators' coefficients flow under
the renormalization group according to
\eqn\coeffeqn{ \mu {d C_i(\mu) \over d \mu} = \sum_j (\gamma^\T)_{ij}
C_j(\mu)}
where $\gamma$ denotes the anomalous dimension matrix which governs how all
operators mix with one another.  $\gamma$ is most readily calculated by
solving the renormalization group equation for 1PI Green's functions
$\Gamma^{(n_\M, n_\B, n_{\bar{\B}})}_i$ with $n_\M$, $n_\B$, $n_{\bar{\B}}$
external meson, baryon and antibaryon lines and insertions of composite
operators labeled by index $i$:
\eqn\RGE{\gamma_{ij} \Gamma^{(n_\M, n_\B, n_{\bar{\B}})}_j
= - \Biggl[ \mu {\partial \over \partial \mu} 
+ \beta {\partial \over \partial \lambda} 
+ \beta^* {\partial \over \partial \lambda^*} 
- n_\M \gamma_\M - (n_\B + n_{\bar{\B}}) \gamma_\B \Biggr] 
\Gamma^{(n_\M, n_\B, n_{\bar{\B}})}_i.}
The meson and baryon anomalous dimensions along with the beta function
entering into this last formula are simply related to the wavefunction
renormalization constants in \wavefunctionconsts:
\eqn\anomdimsandbetafunc{\eqalign{
\gamma_\M &= {1 \over 2} {\mu\over\ZM} {d \ZM \over d\mu} = 
{|\lambda|^2 \over \Nf} + O(|\lambda|^4) \cr
\gamma_\B &= {1 \over 2} {\mu\over\ZB} {d \ZB \over d\mu} = 
|\lambda|^2 + O(|\lambda|^4) \cr
\beta &= \mu {d \lambda \over d\mu} = \lambda \Bigl[ -{\e \over 2} +
\gamma_\M + 2 \gamma_\B \Bigr] = \lambda \Bigl[ -{\e \over 2} + \bigl(2+{1
\over \Nf} \bigr) |\lambda|^2 + O(|\lambda|^4) \Bigr].}}

	The basic forms for one-loop supergraphs with a single insertion of
an $O$, $P$ or $R$ operator which contribute to 2, 3 or 4 point Green's
functions are illustrated in figure~3.  After a long but straightforward
calculation of these diagrams' logarithmically divergent terms, we find the
leading order anomalous dimension matrix takes the block form
\eqn\anomdimmatrix{\gamma \equiv \gammahat |\lambda|^2 = 
\pmatrix{\gamma_{OO} & 0_{3 \times 3} & 0_{3 \times 9} \cr
\gamma_{PO} & 0_{3 \times 3} & 0_{3 \times 9} \cr
0_{9 \times 3} & \gamma_{RP} & \gamma_{RR} \cr} |\lambda|^2 +
O\bigl(|\lambda|^4 \bigr)}
where
\eqn\gammaOOandPO{
\gamma_{OO} = \bordermatrix{&O_1 & O_2 & \bar{O_2} \cr
O_1 & 2/\Nf & 2 & 2 \cr
O_2 & 2/\Nf & 2 & 2 \cr
\bar{O_2} & 2/\Nf & 2 & 2 \cr}
\qquad\qquad
\gamma_{PO} = \bordermatrix{&O_1 & O_2 & \bar{O_2} \cr
P_1 & 4/\Nf &  &  \cr
P_2 &       & 4 &  \cr
\bar{P_2} & & & 4 \cr}}
and $\gamma_{RP} \oplus \gamma_{RR} =$
\def\fouroverNf{{{4 \over \Nf}}}
\def\fouroverNfsq{{{4 \over \Nf^2}}}
\def\twooverNf{{{2 \over \Nf}}}
\def\twooverNfsq{{{2 \over \Nf^2}}}
\eqn\gammaRR{\bordermatrix{&P_1 &P_2 & \bar{P_2} &
R_1 & R_2 & R_3 & R_4 & \bar{R_4} & R_5 & \bar{R_5} & R_6 & \bar{R_6} \cr
R_1 & 0 & 0 & 0 & -4+\twooverNf & 0 & 0 & 0 & 0 &  4 & 4 & {\fouroverNfsq}
& {\fouroverNfsq} \cr
R_2 & 0 & 0 & 0 & 0 & -4+\twooverNf & 0 & 0 & 0 &  4 & 4 & 4 & 4 \cr
R_3 & -{\twooverNf} & 0 & 0 & 0 & 0 & \twooverNf & 2 & 2 & 2 & 2 & 0 & 0 \cr
R_4 & 0 & 0 & 0 & 0 & 0 & 4+\fouroverNf & -\twooverNf & 0 & 4 & 0 &
{\fouroverNf} & 0 \cr
\bar{R_4} & 0 & 0 & 0 & 0 & 0 & 4+\fouroverNf & 0 & -{\twooverNf} & 0 & 4 & 0 
& \fouroverNf  \cr
R_5 & 0 & 0 & -{\twooverNfsq} & {\twooverNf} & 0 & {\twooverNf} &
 \twooverNf & 0  & -2 & 2 & {\fouroverNfsq} & 0 \cr
\bar{R_5} & 0 & -{\twooverNfsq} & 0 & {\twooverNf} & 0 & \twooverNf & 0 & 
\twooverNf & 2 & -2 & 0 & \fouroverNfsq \cr
R_6 & 0 & 0 & -2 & 0 & {\twooverNf} &  \twooverNf & 2 & 0 & 4 & 2 & -2 & 0
\cr
\bar{R_6} & 0 & -2 & 0 & 0 & {\twooverNf} & \twooverNf & 0 & 2 & 2 & 4 & 0
& -2 \cr}.}
The eigenvalues of $\gammahat$ are complicated, unenlightening functions of
$\Nf$.  But with a symbolic manipulator, one may readily check that they
are real for $\Nf \ge 4$ and approach the limiting set $(-4_4, -2_2, 0_7,
4_2)$ with subscripts indicating eigenvalue multiplicities as $\Nf \to
\infty$.

	After decomposing the anomalous dimension matrix as $\gamma = S D
S^{-1} |\lambda|^2 + O(|\lambda|^4)$ where diagonal matrix $D$ contains the
eigenvalues of $\gammahat$ while matrix $S$ holds its eigenvectors, we can
explicitly solve the differential equation \coeffeqn\ for the dimension-4
operators' coefficients:
\eqn\coeffeqnsoln{C_i(\mu) = \sum_{j,k} (S^{-1T})_{ij}\> \Bigl[ {
|\lambda(\mu)|^2 \over |\lambda(|\Lambda|)|^2} \Bigr]^{D_j \over 4+2/\Nf}
 \> (S^T)_{jk} \, C_k(|\Lambda|).}
Although the coefficients of certain linear combinations of $O$, $P$ and
$R$ corresponding to negative eigenvalues of $\gamma$ are logarithmically
enhanced as $\mu \to 0$, these \Kahler\ potential operators are still
irrelevant at long distance scales provided $|\lambda|^2$ is small and
perturbation theory is valid.
So regardless of the signs of nonrenormalizable operator anomalous
dimension eigenvalues, SUSY QCD with $\Nf=\Nc+1$ flavors flows to a free
field theory in the far infrared.

	With only our crude $O(1)$ estimates for Wilson coefficients at the
matching scale $\mu \simeq |\Lambda|$ between the gauge theory and sigma
model, the practical utility of the renormalization group results encoded
into \coeffeqnsoln\ is small at present.  But if precise matching
conditions can someday be determined, the values for \Kahler\ coefficients
at all energies below the confinement scale will then be fixed as well.

\newsec{Flat direction deformations}

	In his original study of SUSY QCD's confining phase, Seiberg
exploited the fact that the theory with $\Nf$ flavors and $\Nc$ colors must
flow to essentially the same model with fewer flavors and colors when one
or more matter fields are given arbitrarily large expectation values and
heavy fields are decoupled \SeibergI.  This consistency requirement plays
an important role in the exact determination of the dynamically generated
superpotential.  As we shall see, the same recursive condition also yields
nontrivial information about the dimensionless coefficients which enter
into the \Kahler\ potential.

\eject

	We will focus upon the relationship between \Kahler\ sector
coefficients at the origin of moduli space and those at the point
\eqn\pointA{\vev{\bareM}=
\pmatrix{0& & &\cr
	 & \ddots & &\cr
	 & & 0 & \cr
	 & & & a^2}
\qquad
\vev{\bareB}=\vev{\bareBbar}=0}
with $a \in \BoldR$.  We have performed a biunitary flavor rotation to
bring the bare meson field into diagonal form and then frozen its $\Nf^{\rm
th}$ component at a classical expectation value whose magnitude greatly
exceeds $\LLdag$.  The large vev for the last meson field may be regarded
as arising from underlying quark and antiquark expectation values which
break the gauge group.  So at this moduli space location, $\Nf=\Nc+1$ SUSY
QCD connects onto the theory with $\Nf-1$ flavors and $\Nc-1$ colors, and
the scales $\Lambda$ and $\LambdaL$ in the upstairs and downstairs
microscopic theories are related by the matching condition
$\Lambda^{2\Nf-3} = a^2 \LambdaL^{2\Nf-5}$.  This classical Higgsing
interpretation makes sense only if $a \gg |\Lambda|$ \tHoofttwo.

	At the point \pointA, the $\Nf^{\rm th}$ diagonal meson fluctuation
is eaten via the Higgs mechanism, while the first $\Nf-1$ baryon
fluctuations are linear in $a$.  The Taylor expansion \Taylorexpansion\ of
the bare \Kahler\ potential therefore looks like
\eqn\KatA{\eqalign{
K &= \KMM(a^2; \Lambda; \Nf) \Tr(\d {\widehat{\bareM}}^\dagger
\d\widehat{\bareM}) + 
a^2  \KBB(a^2; \Lambda; \Nf) \bigl[
{\d \widehat{\bareB}  \d\widehat{\bareB}}^\dagger +
\d{\widehat{\bareBbar}}^\dagger \d \widehat{\bareBbar} \bigr] \cr
& \qquad + \KBB(a^2; \Lambda; \Nf) \bigl[ \d \BzeroNf \d {\bareB^\Nf}^\dagger
 + \d {\BbarzeroNfdagger} \d \bareBbar^\Nf \bigr] + \cdots \cr}}
where hatted fields signify massless hadrons which survive in the low
energy description of $(\Nf-1)=(\Nc-1)+1$ SUSY QCD.  The bare
superpotential can similarly be rewritten in terms of downstairs theory
hadrons:
\eqn\WatA{W = \lambda_0 \Bigl[ {\d \widehat{\bareB} \, \d \widehat{\bareM}
\, \d \widehat{\bareBbar} - \det \d \widehat{\bareM} \over \LambdaL^{2\Nf-5}}
\Bigr] + {\lambda_0 a^2 \over \Lambda^{2\Nf-3}} \d \BzeroNf \d \bareBbar^\Nf.}
After using the superfield equations of motion
\eqn\EOM{\eqalign{
{\Dbar^2 \over 4} {\d K \over \d(\d \BzeroNf)} - {\d W \over \d(\d
\BzeroNf)} &= 0 \cr {D^2 \over 4} {\d K \over \d(\d \BbarzeroNfdagger)} -
{\d \bar{W} \over \d(\d \BbarzeroNfdagger)} &= 0 \cr}}
to integrate out the massive $\d \BzeroNf$ and $\d \BbarzeroNf$
fluctuations and applying superspace identities (A.1) and (A.2), we find
that the kinetic terms for the $\Nf^{\rm th}$ baryon and antibaryon in
\KatA\ are precisely canceled.  The \Kahler\ potential and superpotential
then reduce to those for $(\Nf-1)=(\Nc-1)+1$ SUSY QCD as $a^2 \to \infty$
provided
\eqn\recursionreln{\eqalign{
\KMM(a^2; \Lambda; \Nf) &\equiv {f({a^2 \over \LLdag}; \Nf) \over \LLdag}
\longrightarrow 
\KMM(0; \LambdaL; \Nf-1) = {f(0; \Nf-1) \over \LLLdag} \cr
\KBB(a^2; \Lambda; \Nf) &\equiv {g({a^2 \over \LLdag}; \Nf) \over 
(\LLdag)^{\Nc-1}} \longrightarrow {1 \over a^2} \KBB(0; \LambdaL; \Nf-1) =
{g(0; \Nf-1) \over a^2 (\LLLdag)^{\Nc-2}}. \cr}}
In this recursion relation, we have factored out inverse powers of the SUSY
QCD scale from the \Kahler\ coefficients which are trivially fixed by
dimensional analysis.  The asymptotic $a^2$ dependence of the dimensionless
$f$ and $g$ functions is then basically set by the scale matching
condition.  Recalling $\ZM \propto \KMM \propto f$ and $\ZB \propto \KBB
\propto g$, we deduce 
\eqn\newwavefuncconsts{\eqalign{
\ZM(\mu; a^2;\Nf) &= \Bigl({a^2 \over \LLdag} \Bigr)^{2 \over 2\Nf-5} 
\Bigl[ 1 + O(|\lambda(\mu)|^2) \Bigr] \cr
\ZB(\mu; a^2;\Nf) &= \Bigl({a^2 \over \LLdag} \Bigr)^{-{1 \over 2\Nf-5}}
\Bigl[ 1 + O(|\lambda(\mu)|^2) \Bigr]. \cr}}
As the $\Nf^{\rm th}$ baryon fluctuations were integrated out using just
classical equations of motion, the $O(|\lambda|^2)$ quantum corrections to
the meson and baryon wavefunction renormalization constants are not
determined by these tree-level arguments.  But they could readily be found
by evaluating a few massive one-loop supergraphs.

	The dependence of other \Kahler\ potential couplings upon moduli
space location may be worked out along similar lines.  In general, they
each vary as $a^2 \to \infty$.  On the other hand, all $a^2$ dependence
cancels out from the superpotential couplings as can be seen in
\Wcouplings.  The renormalized $\lambda(\mu)$ expansion parameter consequently
remains uniform throughout the entire moduli space like its bare
$\lambda_0$ progenitor.

\newsec{Conclusion}

	The analysis of the low energy \Kahler\ potential in $\Nf=\Nc+1$
SUSY QCD which we have presented in this article represents only a modest
first step, and many extensions of this work would be interesting to
pursue.  In particular, it would clearly be valuable to determine how
asymptotic limits restrict \Kahler\ sector coefficients at the origin of
moduli space.  As Seiberg emphasized in his study of SUSY QCD's confining
phase superpotential, recovering classical results in weakly coupled
regions of moduli space represents a key constraint which must be satisfied
by the correct quantum description of the low energy theory \SeibergI.  It
may be possible to determine, for instance, how the coefficients of
\Kahler\ potential terms at the moduli space origin must be adjusted so
that $K \to \sqrt{M_0^\dagger M_0}$ along the flat direction $\vev{\det
\bareM} \ne 0$, $\vev{\bareB} = \vev{\bareBbar}=0$ within the classical
domain where the gauge group is completely broken.  Such partial power
series information might provide nontrivial insight into the full
functional form for $K$.

	Other avenues would also be worth exploring.  For example, it
should be straightforward to derive low energy \Kahler\ potentials in SUSY
QCD with $\Nf < \Nc+1$ quark flavors by giving mass to matter fields and
integrating them out from \Kfinal\ and \Wfinal.  The results ought to be
simpler than those which we have found here, for $\Nf < \Nc+1$ SUSY QCD
contains fewer types of hadrons than the theory with $\Nc+1$ flavors.

	Finally, the basic ideas underlying this work can be applied to
investigate \Kahler\ sectors in many other confining supersymmetric gauge
theories.  It would be especially interesting to examine scattering
processes in models with intricate quantum superpotentials.  As we have
seen, leading order kinetic operator coefficients are systematically
calculable in theories with one or more renormalizable superpotential
interaction terms.  So exact moduli space information encoded into
dynamically generated superpotentials should not be completely obscured in
scattering calculations by \Kahler\ potential uncertainties.  Instead, one
may be able to uncover relations among low order hadron scattering
amplitudes.  Such findings would represent a novel departure from vacuum
structure analysis.

\bigskip\bigskip
\centerline{{\bf Acknowledgments}}
\bigskip

	It is a pleasure to thank Marc Grisaru for his many patient and
enlightening tutorials on supergraph perturbation theory.  I am also
grateful to Howard Georgi, Per Kraus, Lisa Randall and Sandip Trivedi for
helpful discussions.  This work was supported by the National Science
Foundation under Grant \#PHY-9218167.

\appendix{A}{Superspace conventions}

	Since the number of different supersymmetry conventions appearing
in the literature is comparable to the total number of supersymmetry
publications, we list here those which we follow in this article along with
some useful superspace identities:

\bigskip\noindent
Spacetime metric signature: $(+,-,-,-)$
\medskip\noindent
Grassmann measures and delta functions:
$$ \eqalignno{
\int d^2 \theta &= - \quarter D^2|_{\theta=\thetabar=0} & (\rm A.1)  \cr
\int d^2 \thetabar &= - \quarter \Dbar^2|_{\theta=\thetabar=0} 
  &  (\rm A.2) \cr
\d^{(4)}(\theta) &= (\theta \theta)(\thetabar \thetabar) &  
  (\rm A.3) \cr} $$
\medskip\noindent
Superspace derivative (anti) commutator relations:
$$ \eqalignno{
\{ D_\a, \Dbar_\bdot \} &= -2 i \sigma^\mu_{a \bdot} \partial_\mu & (\rm A.4) \cr
[ D^2, \Dbar^2] &= 8 i (\Dbar \sigmabar^\mu D)\partial_\mu - 16 \boxderiv &
  (\rm A.5) \cr [ \Dbar^2, D^2] &= 8 i (D \sigma^\mu \Dbar) \partial_\mu -
  16 \boxderiv & (\rm A.6) \cr} $$
\medskip\noindent
$D$-algebra identities:
$$ \eqalignno{
\Dbar\sigmabar^\mu D + D \sigma^\mu \Dbar &= -4 i \partial^\mu & 
  (\rm A.7) \cr
\d^{(4)}(\theta_x-\theta_y) D^2 \Dbar^2 \d^{(4)}(\theta_x-\theta_y) &=
16 \d^{(4)}(\theta_x-\theta_y) & (\rm A.8) \cr} $$
\medskip\noindent
Chiral superfield functional derivatives:
$$ \eqalignno{
& {\d \Phi(x,\theta) \over \d \Phi(x',\theta')} = -{\Dbar^2 \over 4} 
\d^{(4)}(x-x') \d^{(4)}(\theta-\theta') & (\rm A.9) \cr
& {\d \Phi^\dagger(x,\thetabar) \over \d \Phi^\dagger(x',\thetabar')} 
= -{D^2 \over 4} \d^{(4)}(x-x') \d^{(4)}(\theta-\theta') & (\rm A.10) \cr} $$ 

\listrefs
\bye